\documentclass[review]{elsarticle}

\usepackage{lineno}
\modulolinenumbers[5]

\usepackage[utf8]{inputenc}
\usepackage[T1]{fontenc}
\usepackage{natbib}
\biboptions{authoryear}
\usepackage{graphicx}
\usepackage[english]{babel}
\usepackage{amssymb}
\usepackage[cmex10]{amsmath}
\usepackage{algorithm}
\usepackage{algpseudocode}
\usepackage{array}
\usepackage[tight,footnotesize]{subfigure}
\usepackage{fixltx2e}
\usepackage{microtype}
\usepackage[binary-units=true]{siunitx}
\usepackage{psfrag}
\usepackage{pst-sigsys}
\usepackage{pstricks}
\usepackage{pgfplots}
\usepackage{pgfkeys}
\usepackage{booktabs}
\usepackage{ae}
\usepackage{url}
\usepackage[nolist,nohyperlinks]{acronym}
\usepackage{hyperref}
\usepackage{siunitx}
\usepackage{etex}
\usepackage{tikz}
\usepackage{pgfkeys}
\usepackage{calc}
\usepackage{pgfplots}
\usetikzlibrary{arrows}
\usepgflibrary{arrows}
\usetikzlibrary{positioning}
\usetikzlibrary{intersections}
\pgfplotsset{compat=newest}
\usetikzlibrary{shapes.geometric}

\makeatletter
\newcommand{\gettikzxy}[3]{%
  \tikz@scan@one@point\pgfutil@firstofone#1\relax
  \edef#2{\the\pgf@x}%
  \edef#3{\the\pgf@y}%
}
\makeatother

\tikzset{
	node style main block/.style={draw, minimum width = 2cm, minimum 
		height=1cm, text height=1.5ex, text depth=.25ex},
	triangle/.style = { regular polygon, regular polygon sides=3, shape border rotate=270,inner sep=1pt}
}

\definecolor{dark_red}{rgb}{0.5, 0.0, 0.0}
\definecolor{dark_blue}{rgb}{0.0, 0.0, 0.6}
\definecolor{light_blue}{rgb}{0.5, 0.5, 1.0}
\definecolor{light_orange}{rgb}{1,0.7490,0.5020}
\definecolor{dark_orange}{rgb}{1,0.5020,0}
\definecolor{light_red}{rgb}{1.0, 0.5, 0.5}
\definecolor{light_green}{rgb}{0.5, 1.0, 0.5}
\definecolor{dark_green}{rgb}{0.0, 0.6, 0.0}
\definecolor{ddark_green}{rgb}{0.0, 0.35, 0.0}
\definecolor{ddark_blue}{rgb}{0.0, 0.0, 0.35}
\definecolor{seahorse}{rgb}{0.839, 0.839, 0.937}
\definecolor{ddark_seahorse}{rgb}{0.739, 0.739, 0.837}
\definecolor{dark_seahorse}{rgb}{0.757, 0.757, 0.909}
\definecolor{g_o_1}{rgb}{0.9725,0.7961,0.6784}
\definecolor{g_o_2}{rgb}{0.9569,0.6902,0.5176}
\definecolor{g_o_3}{rgb}{0.9294,0.4902,0.1922}
\definecolor{g_b_1}{rgb}{0.7412,0.8431,0.9333}
\definecolor{g_b_2}{rgb}{0.6078,0.7608,0.9020}
\definecolor{g_b_3}{rgb}{0.3569,0.6078,0.8353}
\definecolor{excel_good_fill}{rgb}{0.7765,0.9373,0.8078}
\definecolor{excel_good_text}{rgb}{0,0.3804,0}
\definecolor{opacity_text}{rgb}{0.4588,0.4588,0.4824}
\newlength\figureheight
\newlength\figurewidth
\newcommand{\lrp}[1]{\ensuremath{\left( #1 \right)}}

\newcommand{\ui}[2]{#1 _{\text{#2}}}

\newcommand{\qin}{\ui{q}{in}}
\newcommand{\qout}{\ui{q}{out}}
\newcommand{\Ts}{\ui{T}{s}}

\newcommand{\diff}[2]{\ensuremath{\frac{\text{d}\ \! #1}{\text{d}\ \! #2}}}
\newcommand{\dift}[1]{\frac{\text{d}\ \! #1}{\text{d}\ \! t}}

\newtheorem{theorem}{Theorem}[section]

\newtheorem{remark}[theorem]{Remark}

\usepackage{natbib}
\setcitestyle{authoryear,round}

\usetikzlibrary{external}
\newcommand{\includetikz}[1]{%
	\tikzifexternalizing{%
		\def\DOIT{1}%
	}{%
		\IfFileExists{#1.pdf}{%
			\includegraphics[scale=1]{#1.pdf}%
			\def\DOIT{0}%
		}{%
			\def\DOIT{1}%
		}%
	}%
	\if1\DOIT
	\tikzsetnextfilename{#1}
	\input{#1.tikz}
	\fi
}

\journal{Acta Chimica Slovaca}
\begin{document}

\begin{frontmatter}

% continue with document
\title{Non-linear Model Predictive Control of Conically Shaped Liquid Storage 
Tanks}

\author{Martin Klau\v co}
\author{\v Lubo\v s \v Cirka}

\address{Slovak University of Technology in Bratislava,\\
  Radlinsk\'eho 9, SK-812 37 Bratislava, Slovak Republic \\
  {\tt\small\{martin.klauco,lubos.cirka\}@stuba.sk}, +421 259 325 
  345}

\begin{abstract}
This paper deals with the analysis and synthesis of a model predictive control 
(MPC) strategy used in connection with level control in conically shaped 
industrial liquid storage tanks. The MPC is based on a dynamical non-linear 
model describing the changes of the liquid level with respect to changes in the 
inlet flow of the liquid. An Euler discretization of the dynamical system is 
exploited to transform the continuous time dynamics to its discrete time 
counterpart, used in the non-linear MPC (NMPC) synthesis. By means of 
a simulation case study will be shown, that NMPC better tracks the changes of 
the liquid level, hence provides increased control performance. This paper also 
compares the traditional approach of optimal control, the linear MPC, with the 
NMPC strategy. 
\end{abstract}

\end{frontmatter}
\linenumbers

\section{Introduction}
\label{sec:introduction}

The model predictive control is a well-established control strategy in chemical 
process control. The main advantages stem from optimally shaping the trajectory 
of manipulated variables with respect to performance criteria and 
technological and safety constraints~\citep{mayne:aut:2000,camacho:2007}. The 
optimal control strategies have been systematically addressed in countless 
scientific works, including time optimal control~\citep{uiam1612}, or standard 
model predictive control~\citep{muske:jpc:2002:ofs,uiam906,uiam1552}. All 
aforementioned works, however, focus on the standardized design of the model 
predictive control, which relies on linear state space models of the controlled 
plant. Such approaches, however, introduce an obstacle, which is called 
"model-mismatch", where the design model in the controller does not match the 
actual process. 

To remedy the situation, researchers focus on non-linear model predictive 
control (NMPC), which improves given control strategies by incorporating the 
non-linear equations capturing the dynamics of the 
system~\citep{Allgwer2004NonlinearMP}. This work focuses on the application of 
such a controller to the most common chemical process, which is the control of 
a level of the liquid inside storage tanks. Specifically, we focus on a 
conically-shaped liquid storage tank.

This paper is organized as follows. First, we introduce the non-linear 
mathematical model of conical tank. Second, we focus on the synthesis of two 
controllers, the linear MPC and the non-linear MPC. Lastly, we compare the 
performance of aforementioned controllers by the means of simulation case study.

\section{Mathematical Modeling of Conically Shaped Tanks}
The dynamical mathematical model of a tank with one inlet stream, denoted as 
$\qin(t)$ and one outlet stream given by $\qout(t)$, is given by a mass balance 
equation of following form
\begin{equation}
	\label{eq:tank_V}
	\qin(t) = \qout(t) + \dift{V(t)},
\end{equation}
where the $V(t)$ stands for the volume of a liquid inside the 
tank. In this work, we consider the level of the liquid inside the tank as a 
process variable, hence we rewrite the model in~\eqref{eq:tank_V} to
\begin{equation}
	\label{eq:tank_HV}
	\qin(t) = \ui{k}{v}\sqrt{h(t)} + 
  \diff{V(t)}{h} \dift{h(t)},
\end{equation}
and we define
\begin{equation}
	\label{eq:tank_F}
	F(h) = \diff{V(t)}{h}.
\end{equation}
For the purpose of performing simulations, we convert the model 
in~\eqref{eq:tank_HV} to a non-linear state space form
\begin{equation}
	\label{eq:tank_H}
	\dift{h(t)}	 = \frac{1}{F(h)}\lrp{\qin(t) - \ui{k}{v}\sqrt{h(t)}}.
\end{equation}
The variable $\ui{k}{v}$ correspond to an output valve coefficient. The valve 
coefficient can be derived from Bernoulli equation, and it represents the 
friction of liquid movement in the outlet pipe~\cite[ch.~2]{fikar:2007:book}.

In this work we consider a controller synthesis, which is based on a discrete 
time model, hence the non-linear system model can obtained by Euler 
discretization of~\eqref{eq:tank_H}. Specifically,
\begin{align}
	\label{eq:euler}
	h(t + \Ts) &= h(t) + \Ts\cdot\lrp{ \frac{1}{F(h)}\lrp{\qin(t) - 
	\ui{k}{v}\sqrt{h(t)}} },
\end{align}
where the variable $\Ts$ represent the sampling time. Even though the Euler 
discretization process can be inexact, it is often used in controller 
design as suggested by~\cite{lavrynczuk:2017:isa}.

\begin{figure}
	\centering
		\resizebox {0.7\textwidth} {!} {
	\begin{tikzpicture}[rotate=270],[transform shape]
	\draw (0,0) arc (-90:90:0.6 and 1.5);% right half of the 
	\draw[dashed] (2,0.5) arc (-90:90:0.4 and 0.99);
	\draw[dashed] (2,0.5) arc (270:90:0.4 and 0.99);
	\draw (0,0) -- (4,1);% bottom line
	\draw (0,4) -- (4,5);
	\draw (0,3) -- (4,2);% top line
	\draw (0,7) -- (4,6);
	\draw (0,4) -- (0,7);
	\draw (4,5) -- (4,6);
	\draw[|-|] (0,8) -- (4,8);
	\draw[|-|] (0,5.5) -- (0,7);
	\draw[|-|] (4,5.5) -- (4,6);
	\node (0) at (4,8.5) {$0$};
	\node (hmax) at (0,8.7) {$h_\mathrm{max}$};
	\node (ht) at (2,8.6) {$h(t)$};
	\node (r1) at (-0.3,6.2) {$R_1$};
	\node (r2) at (4.3,5.75) {$R_2$};
	\node (rf) at (1.8,6) {$r_{\mathrm{f}}$};
	\draw[dotted] (0,5.5) -- (4,5.5);
	\draw[densely dashed] (2,2.5) -- (2,8);
	\draw[densely dashed] (0,3) -- (0,4);
	\draw[densely dashed] (4,2) -- (4,5);
	\draw (0,0) arc (270:90:0.6 and 1.5);% left half of the left ellipse
	\draw (4,1.5) ellipse (0.166 and 0.5);% right ellipse
	\draw[-triangle 45] (-2,1.5) -- (-0.7,1.5);
	\draw (-2,0) -- (-2,1.5);
	\draw (4.3,1.5) -- (5.5,1.5);
	\draw[-triangle 45] (5.5,1.5) -- (5.5,3);
	\node (q0) at (-2.4, 0) {$q_{\mathrm{in}}(t)$};
	\node (q1) at (5.9, 3) {$q_{\mathrm{out}}(t)$};
	\end{tikzpicture}
}
	\label{fig:tank:conical}
	\caption{Illustration of the conically-shaped tank.}
\end{figure}
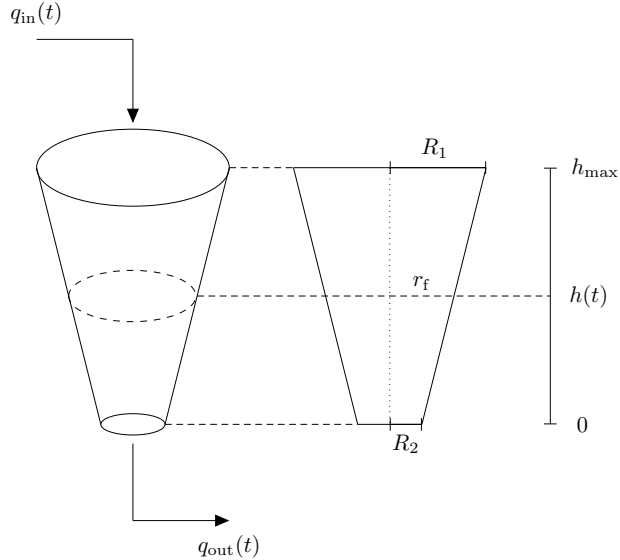

We consider an inverted frustum of a right cone as an open conical tank 
process. The geometrical representation of the conical tank is shown in the 
Fig.~\ref{fig:tank:conical}. The model of such a process is based on findings 
by~\cite{king:2010:book}, and it is derived by expressing the volume of the 
frustum as a function of the level of the liquid. The 
tank is characterized by variables $R_1$, $R_2$, which are radii of the bottom 
and upper base, respectively and by the height $\ui{h}{max}$ 
(cf.~Fig.\ref{fig:tank:conical}). The volume of the liquid inside the frustum 
is given by
\begin{equation}
	\label{eq:frustum_V}
	\ui{V}{f}(h(t)) = \frac{\pi h(t)}{3}\lrp{\ui{r}{f}^2(h(t)) + 
	R_2\ui{r}{f}(h(t)) + R_2^2},
\end{equation}
where the variable $\ui{r}{f}(h(t))$ is the radius of a disc representing the 
surface of the liquid at level $h(t)$. The radius $\ui{r}{f}(h(t))$ is explicit 
function of the liquid level, expressed as
\begin{equation}
	\label{eq:frustum_rf}
	\ui{r}{f}(h(t)) = R_2 + \frac{R_1 - R_2}{\ui{h}{max}}h(t).
\end{equation}
By substituting the expression in~\eqref{eq:frustum_rf} to~\eqref{eq:frustum_V} 
we obtain
\begin{equation}
	\label{eq:frustum_h}
	\ui{V}{f}(h(t)) = \frac{\pi h(t)}{3}\lrp{ 3R_2^2 + 3R_2\frac{R_1 - 
	R_2}{\ui{h}{max}}h(t) + \lrp{\frac{R_1 - R_2}{\ui{h}{max}}}^2h^2(t)}.
\end{equation}
Next, we combine the expression for the volume in~\eqref{eq:frustum_h} and the 
general mass balance model in~\eqref{eq:tank_HV}, which results in
\begin{equation}
	\label{eq:conical}
	\qin(t) = \ui{k}{v}\sqrt{h(t)} + \pi \lrp{R_2 + 	
	h(t)\frac{R_1-R_2}{\ui{h}{max}} }^2\dift{h(t)}.
\end{equation}

Symbols, physical quantities and parameters are reported in the 
table~\ref{tab:params}. The non-linear mathematical model reported 
in~\eqref{eq:conical} is used in the synthesis of the NMPC strategy, addressed 
in the next section.

\begin{table}
	\centering
	\caption{Parameters of the conical tank system and quantities related to 
	system dynamics.}
	\label{tab:params}
	\begin{tabular}{lcl}
		\toprule
		Physical quantity & Symbol & Value \\
		\midrule
		Height steady state & $\ui{h}{L}$ & \SI{0.4000}{\metre} \\
		Inlet steady state & $\ui{q}{in,L}$ &
		\SI{0.0474}{\metre\cubed\per\second} \\
		Valve coefficient & $\ui{k}{v}$ & ${0.0750\;\text{m}^{2.5}\text{s}^{-1}}$ \\
		Maximum height & $\ui{h}{max}$ & \SI{2.0000}{\metre} \\
		Upper radius  & $R_1$ &\SI{1.0000}{\metre} \\
		Bottom radius  & $R_2$ & \SI{0.4000}{\metre} \\
		Minimum flow  & $\ui{q}{in,min}$ & \SI{0.0000}{\metre\cubed\per\second} \\
		Maximum flow  & $\ui{q}{in,max}$ & \SI{0.1000}{\metre\cubed\per\second} \\
%		\midrule
		Sampling time  & $\Ts$ & \SI{2.0000}{\second} \\
		\bottomrule
	\end{tabular}
\end{table}

\section{Synthesis of Controllers}
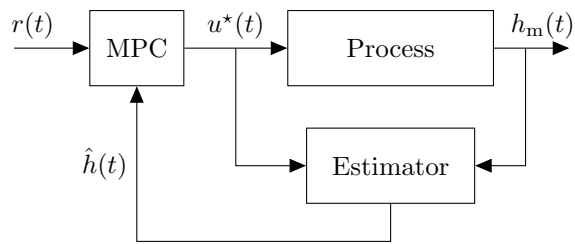
\begin{figure}
	\centering
	\begin{tikzpicture}
  \node[draw, minimum width = 2.5cm, text width = 2.5cm, align = 
  center, minimum height = 1.0cm] (process) {Process};
  
  \node [draw, left = 2cm of process, minimum width = 1.25cm, minimum 
  height = 1.0cm, anchor = center] (mpc) {MPC};

  \draw [-triangle 45] (mpc) -- node [above] (sig_u) {$u^{\star}(t)$} (process);
  
  % kalman filter
  \node [draw, minimum height = 1cm, minimum width = 2cm, text width = 
  2cm, below = 0.55cm of process, align = center] (kf) {Estimator};

  % from process to kf
  \draw [-triangle 45] (process.0) ++(0.42cm, 0) |- (kf.0);

  % from mpc to kf
  \draw [-triangle 45] (sig_u.270) |- (kf.180);

  % from kf to mpc
  \draw [-triangle 45] (kf.270) -- ++(0cm, -0.5cm) -|  node [near end, left] {$\hat{h}(t)$} (mpc.270);
  
  % this is used to center the block diagram
  % \node [left = 1cm of sum.180] {~};
  
  % measured output
  \draw [-triangle 45] (process.0) -- node [above, xshift = 
  +0.15cm] {$h_{\text{m}}(t)$}   ++(1.0cm, 0);
  
  \node[left = 1cm of mpc] (ref) {~};
  \draw[-triangle 45] (ref) -- node [near start, above] {$r(t)$} 
  (mpc.180);
  
\end{tikzpicture}
	\caption{General model predictive control strategy scheme. The $r(t)$ stands 
	for the reference signal, i.e., the desired level of the liquid, next the 
	$u^{\star}(t)$ is the optimal control action, i.e., the inlet flow of liquid. 
	The actual measurement of the liquid level is depicted by $\ui{h}{m}(t)$, 
	while the estimate of the level is denoted by $\hat{h}(t)$.}
	\label{fig:mpcScheme}
\end{figure}

In this work, we consider the synthesis of the non-linear model predictive 
control strategy, which exploits the non-linear nature of the dynamical model. 
In order to demonstrate the benefits of the non-linear controller, we compare 
this approach with the standardized linear version of the MPC. Both of these 
controllers are implemented in scheme depicted on the Fig.~\ref{fig:mpcScheme}.

The closed-loop control is realized also with an estimator, which purpose is to 
estimate possible mismatch between the design model and the actual process. 
Such a control strategy has been adopted from works 
by~\cite{rawlings:book:2009} and~\cite{muske:acc:1997:ofs}.

The synthesis and implementation of model predictive control follow
the principles receding horizon policy established by~\cite{mayne:aut:2000}. It 
optimizes control actions over a prediction horizon $N$ based on predictions of 
the future trajectory of the process variable. 

Specifically, the non-linear model predictive controller is casted as an 
optimization problem with a quadratic cost function and nonlinear equality 
constraints, 
\begin{subequations}
	\label{eq:nmpc}
	\begin{align}
		\min _{u_0, \ldots, u_{N-1}} \; & \; \sum_{k = 0}^{N-1} \lrp{   
		 ||(x_k - r_k)||_{\ui{Q}{x}}^{2} - ||(u_k - u_{k-1})||_{\ui{Q}{u}}^{2}
		 } 
		\label{eq:nmpc:obj}\\
		\text{s.t.} \; & \; x_{k+1} = x_k + \Ts\cdot h(x_k, u_k), 
		\label{eq:nmpc:xk}\\
		 \; & \; x_k \in [\ui{h}{min}, \; \ui{h}{max}], \label{eq:nmpc:x}\\
		 \; & \; u_k \in [\ui{q}{in,min}, \; \ui{q}{in,max}], \label{eq:nmpc:u}\\
		 \; & \; (u_k - u_{k-1}) \in [\Delta \ui{q}{in,min}, \; \Delta 
		 \ui{q}{in,max}], 
		 \label{eq:nmpc:du}\\
		 \; & \; x_0 = h(t),\; u_{-1} = u(t-\Ts). \label{eq:nmpc:x0}
	\end{align}
\end{subequations}
The objective function~\eqref{eq:nmpc:obj} penalizes the difference between 
prediction of the liquid level $x_k$ and height reference $r_k$, followed by a 
second term which penalizes the increments of control actions. Such a structure 
of the objective function enforces offset-free control 
performance~\citep{muske:jpc:2002:ofs}. Note, that the term 
$||z||_M^2 = z^\intercal Mz$ represents a squared Euclidean norm. The 
prediction equation~\eqref{eq:nmpc:xk} is represented by the non-linear 
dynamical model from~\eqref{eq:euler}. Constraints~\eqref{eq:nmpc:x} 
and~\eqref{eq:nmpc:u} ensure, that technological limits on the process 
variable, and on the manipulated variable are satisfied. Namely, the 
constraint~\eqref{eq:nmpc:x} represent physical dimension of the tank, the 
constraint~\eqref{eq:nmpc:u} defines the range of inlet flow, while 
the equation~\eqref{eq:nmpc:du} bounds how fast the inlet flow can change, 
i.e., how fast can the control valve by open or closed. Lastly, the 
optimization problem is initialized by the current measurement of the height 
and by previous control action, as in~\eqref{eq:nmpc:x0}, and
constraints~\eqref{eq:nmpc:xk}-\eqref{eq:nmpc:du} are enforced for $k = 0, 
\ldots, N-1$.

The optimization problem given by~\eqref{eq:nmpc} can be solved by 
off-the-shelve tools like \textit{fmincon} in Matlab, which exploits procedures 
like interior-point method or trust-region method~\citep{nocedal:book:2006}.

The linear version of the MPC has the same form as the non-linear version given 
by~\eqref{eq:nmpc}, except the constraint in~\eqref{eq:nmpc:xk} which represent 
the prediction equation. Here, the non-linear dynamical equation is linearized 
by Taylor first-order expansion around an operating point (cf. 
Remark~\ref{rem:ss}) denoted as $(\ui{h}{L}, \ui{q}{in,L})$. The resulting 
prediction equation has the form of a linear state space model, which is 
subsequently discretized by a sampling time $\Ts$, specifically
\begin{equation}
	x(t + \Ts) = Ax(t) + Bu(t),
\end{equation}
where the state vector $x(t)$ and control input $u(t)$ is define as a deviation 
from respective steady states values. The linear MPC is then casted as 
quadratic optimization problem (QP) with linear constraints. This QP problem 
can be solved by \textit{quadprog} function in Matlab, or by GUROBI solver.  
Note, that the synthesis of individual controllers is a general 
procedure, however, we used parameters from the table~\ref{tab:params} to 
construct the optimization problems.

\begin{remark}
	\label{rem:ss}
	The operating point, often called a steady state, 
	can be explicitly calculated from the non-linear mode in~\eqref{eq:tank_H} by 
	solving $\frac{1}{F(\ui{h}{L})}\lrp{\ui{q}{in,L} - \ui{k}{v}\sqrt{\ui{h}{L}}} 
	= 0$. Note, that the choice of operating point affects the performance of 
	linear-based control strategies. Note, that the linearisation point should be 
	chosen with respect to technological properties of the plant.
\end{remark}

\section{Comparisons and Results}
\label{sec:cmp}

\begin{figure}
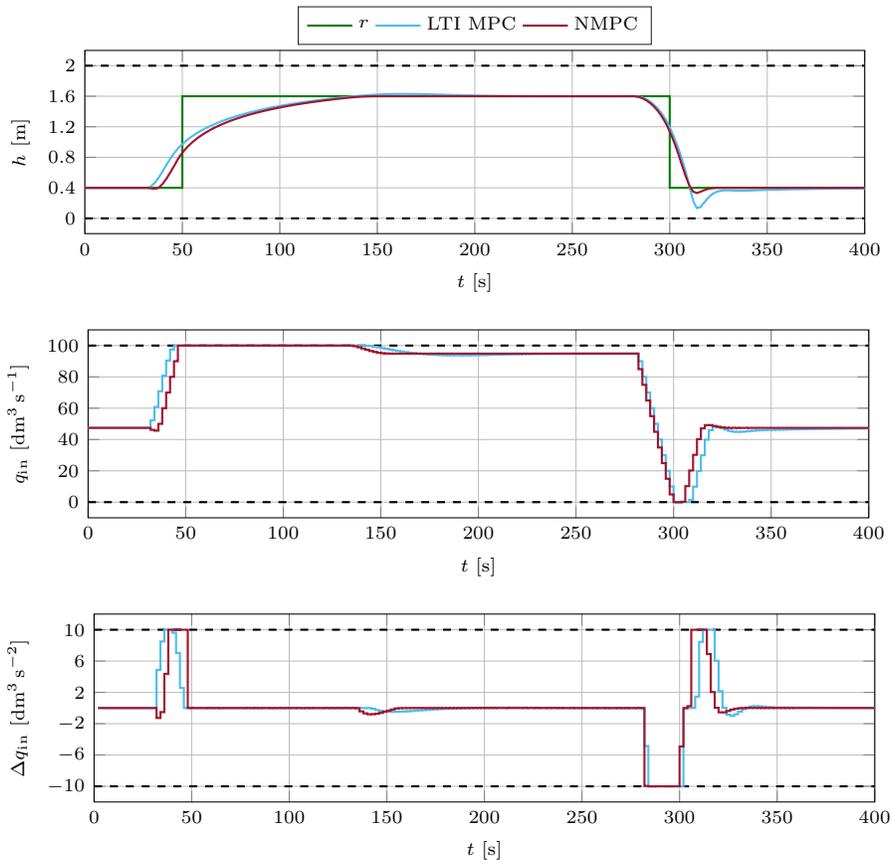

	\centering
	\setlength\figureheight{2.5cm}
	\setlength\figurewidth{0.855\linewidth}
	\subfigure{\input{images/output.tikz}}\\
	\setlength\figureheight{2.5cm}
	\setlength\figurewidth{0.9\linewidth}
	\subfigure{% This file was created by matlab2tikz.
%
\definecolor{mycolor1}{rgb}{0.30196,0.74510,0.93333}%
\definecolor{mycolor2}{rgb}{0.63529,0.07843,0.18431}%
\begin{tikzpicture}

\begin{axis}[%
width=0.951\figurewidth,
height=\figureheight,
at={(0\figurewidth,0\figureheight)},
scale only axis,
xmin=0,
xmax=400,
xtick={  0,  50, 100, 150, 200, 250, 300, 350, 400},
xlabel style={font=\color{white!15!black}},
xlabel={$t~[\SI{}{\second}]$},
ymin=-10,
ymax=110,
ytick={  0,  20,  40,  60,  80, 100},
ylabel style={font=\color{white!15!black}},
ylabel={$\ui{q}{in}~[\SI{}{\deci\metre\cubed\per\second}]$},
axis background/.style={fill=white},
xmajorgrids,
ymajorgrids,
ylabel style={font=\scriptsize},xlabel style={font=\scriptsize},xticklabel style={font=\scriptsize, /pgf/number format/.cd, 1000 sep={}},yticklabel style={font=\scriptsize},legend style={font=\scriptsize}
]
\addplot[const plot, color=black, dashed, line width=0.8pt, forget plot] table[row sep=crcr] {%
0	0\\
2	0\\
4	0\\
6	0\\
8	0\\
10	0\\
12	0\\
14	0\\
16	0\\
18	0\\
20	0\\
22	0\\
24	0\\
26	0\\
28	0\\
30	0\\
32	0\\
34	0\\
36	0\\
38	0\\
40	0\\
42	0\\
44	0\\
46	0\\
48	0\\
50	0\\
52	0\\
54	0\\
56	0\\
58	0\\
60	0\\
62	0\\
64	0\\
66	0\\
68	0\\
70	0\\
72	0\\
74	0\\
76	0\\
78	0\\
80	0\\
82	0\\
84	0\\
86	0\\
88	0\\
90	0\\
92	0\\
94	0\\
96	0\\
98	0\\
100	0\\
102	0\\
104	0\\
106	0\\
108	0\\
110	0\\
112	0\\
114	0\\
116	0\\
118	0\\
120	0\\
122	0\\
124	0\\
126	0\\
128	0\\
130	0\\
132	0\\
134	0\\
136	0\\
138	0\\
140	0\\
142	0\\
144	0\\
146	0\\
148	0\\
150	0\\
152	0\\
154	0\\
156	0\\
158	0\\
160	0\\
162	0\\
164	0\\
166	0\\
168	0\\
170	0\\
172	0\\
174	0\\
176	0\\
178	0\\
180	0\\
182	0\\
184	0\\
186	0\\
188	0\\
190	0\\
192	0\\
194	0\\
196	0\\
198	0\\
200	0\\
202	0\\
204	0\\
206	0\\
208	0\\
210	0\\
212	0\\
214	0\\
216	0\\
218	0\\
220	0\\
222	0\\
224	0\\
226	0\\
228	0\\
230	0\\
232	0\\
234	0\\
236	0\\
238	0\\
240	0\\
242	0\\
244	0\\
246	0\\
248	0\\
250	0\\
252	0\\
254	0\\
256	0\\
258	0\\
260	0\\
262	0\\
264	0\\
266	0\\
268	0\\
270	0\\
272	0\\
274	0\\
276	0\\
278	0\\
280	0\\
282	0\\
284	0\\
286	0\\
288	0\\
290	0\\
292	0\\
294	0\\
296	0\\
298	0\\
300	0\\
302	0\\
304	0\\
306	0\\
308	0\\
310	0\\
312	0\\
314	0\\
316	0\\
318	0\\
320	0\\
322	0\\
324	0\\
326	0\\
328	0\\
330	0\\
332	0\\
334	0\\
336	0\\
338	0\\
340	0\\
342	0\\
344	0\\
346	0\\
348	0\\
350	0\\
352	0\\
354	0\\
356	0\\
358	0\\
360	0\\
362	0\\
364	0\\
366	0\\
368	0\\
370	0\\
372	0\\
374	0\\
376	0\\
378	0\\
380	0\\
382	0\\
384	0\\
386	0\\
388	0\\
390	0\\
392	0\\
394	0\\
396	0\\
398	0\\
400	0\\
};
\addplot[const plot, color=black, dashed, line width=0.8pt, forget plot] table[row sep=crcr] {%
0	100\\
2	100\\
4	100\\
6	100\\
8	100\\
10	100\\
12	100\\
14	100\\
16	100\\
18	100\\
20	100\\
22	100\\
24	100\\
26	100\\
28	100\\
30	100\\
32	100\\
34	100\\
36	100\\
38	100\\
40	100\\
42	100\\
44	100\\
46	100\\
48	100\\
50	100\\
52	100\\
54	100\\
56	100\\
58	100\\
60	100\\
62	100\\
64	100\\
66	100\\
68	100\\
70	100\\
72	100\\
74	100\\
76	100\\
78	100\\
80	100\\
82	100\\
84	100\\
86	100\\
88	100\\
90	100\\
92	100\\
94	100\\
96	100\\
98	100\\
100	100\\
102	100\\
104	100\\
106	100\\
108	100\\
110	100\\
112	100\\
114	100\\
116	100\\
118	100\\
120	100\\
122	100\\
124	100\\
126	100\\
128	100\\
130	100\\
132	100\\
134	100\\
136	100\\
138	100\\
140	100\\
142	100\\
144	100\\
146	100\\
148	100\\
150	100\\
152	100\\
154	100\\
156	100\\
158	100\\
160	100\\
162	100\\
164	100\\
166	100\\
168	100\\
170	100\\
172	100\\
174	100\\
176	100\\
178	100\\
180	100\\
182	100\\
184	100\\
186	100\\
188	100\\
190	100\\
192	100\\
194	100\\
196	100\\
198	100\\
200	100\\
202	100\\
204	100\\
206	100\\
208	100\\
210	100\\
212	100\\
214	100\\
216	100\\
218	100\\
220	100\\
222	100\\
224	100\\
226	100\\
228	100\\
230	100\\
232	100\\
234	100\\
236	100\\
238	100\\
240	100\\
242	100\\
244	100\\
246	100\\
248	100\\
250	100\\
252	100\\
254	100\\
256	100\\
258	100\\
260	100\\
262	100\\
264	100\\
266	100\\
268	100\\
270	100\\
272	100\\
274	100\\
276	100\\
278	100\\
280	100\\
282	100\\
284	100\\
286	100\\
288	100\\
290	100\\
292	100\\
294	100\\
296	100\\
298	100\\
300	100\\
302	100\\
304	100\\
306	100\\
308	100\\
310	100\\
312	100\\
314	100\\
316	100\\
318	100\\
320	100\\
322	100\\
324	100\\
326	100\\
328	100\\
330	100\\
332	100\\
334	100\\
336	100\\
338	100\\
340	100\\
342	100\\
344	100\\
346	100\\
348	100\\
350	100\\
352	100\\
354	100\\
356	100\\
358	100\\
360	100\\
362	100\\
364	100\\
366	100\\
368	100\\
370	100\\
372	100\\
374	100\\
376	100\\
378	100\\
380	100\\
382	100\\
384	100\\
386	100\\
388	100\\
390	100\\
392	100\\
394	100\\
396	100\\
398	100\\
400	100\\
};
\addplot[const plot, color=mycolor1, line width=0.8pt, forget plot] table[row sep=crcr] {%
0	47.43\\
2	47.43\\
4	47.43\\
6	47.43\\
8	47.43\\
10	47.43\\
12	47.43\\
14	47.43\\
16	47.43\\
18	47.43\\
20	47.43\\
22	47.43\\
24	47.43\\
26	47.43\\
28	47.43\\
30	47.43\\
32	52.3\\
34	60.79\\
36	70.79\\
38	80.79\\
40	90.41\\
42	97.43\\
44	100\\
46	100\\
48	100\\
50	100\\
52	100\\
54	100\\
56	100\\
58	100\\
60	100\\
62	100\\
64	100\\
66	100\\
68	100\\
70	100\\
72	100\\
74	100\\
76	100\\
78	100\\
80	100\\
82	100\\
84	100\\
86	100\\
88	100\\
90	100\\
92	100\\
94	100\\
96	100\\
98	100\\
100	100\\
102	100\\
104	100\\
106	100\\
108	100\\
110	100\\
112	100\\
114	100\\
116	100\\
118	100\\
120	100\\
122	100\\
124	100\\
126	100\\
128	100\\
130	100\\
132	100\\
134	100\\
136	100\\
138	100\\
140	100\\
142	99.91\\
144	99.66\\
146	99.3\\
148	98.87\\
150	98.4\\
152	97.92\\
154	97.44\\
156	96.98\\
158	96.54\\
160	96.13\\
162	95.76\\
164	95.42\\
166	95.11\\
168	94.85\\
170	94.61\\
172	94.41\\
174	94.24\\
176	94.1\\
178	93.99\\
180	93.9\\
182	93.84\\
184	93.79\\
186	93.77\\
188	93.76\\
190	93.76\\
192	93.77\\
194	93.8\\
196	93.83\\
198	93.87\\
200	93.92\\
202	93.97\\
204	94.02\\
206	94.07\\
208	94.13\\
210	94.19\\
212	94.24\\
214	94.29\\
216	94.35\\
218	94.4\\
220	94.45\\
222	94.49\\
224	94.54\\
226	94.58\\
228	94.62\\
230	94.65\\
232	94.68\\
234	94.71\\
236	94.74\\
238	94.76\\
240	94.79\\
242	94.81\\
244	94.82\\
246	94.84\\
248	94.85\\
250	94.86\\
252	94.87\\
254	94.88\\
256	94.89\\
258	94.89\\
260	94.9\\
262	94.9\\
264	94.9\\
266	94.91\\
268	94.91\\
270	94.91\\
272	94.91\\
274	94.91\\
276	94.91\\
278	94.9\\
280	94.9\\
282	90.05\\
284	80.05\\
286	70.05\\
288	60.05\\
290	50.05\\
292	40.05\\
294	30.05\\
296	20.05\\
298	10.05\\
300	0.04895\\
302	1.669e-10\\
304	4.321e-10\\
306	9.263e-12\\
308	1.456\\
310	9.9\\
312	19.9\\
314	29.9\\
316	39.9\\
318	45.94\\
320	48.09\\
322	48.12\\
324	47.27\\
326	46.26\\
328	45.46\\
330	45\\
332	44.86\\
334	44.94\\
336	45.14\\
338	45.38\\
340	45.61\\
342	45.79\\
344	45.92\\
346	46.02\\
348	46.09\\
350	46.15\\
352	46.2\\
354	46.26\\
356	46.31\\
358	46.37\\
360	46.42\\
362	46.48\\
364	46.53\\
366	46.58\\
368	46.63\\
370	46.67\\
372	46.71\\
374	46.75\\
376	46.79\\
378	46.82\\
380	46.85\\
382	46.88\\
384	46.91\\
386	46.94\\
388	46.97\\
390	46.99\\
392	47.01\\
394	47.04\\
396	47.06\\
398	47.08\\
400	47.1\\
};
\addplot[const plot, color=mycolor2, line width=0.8pt, forget plot] table[row sep=crcr] {%
0	47.43\\
2	47.43\\
4	47.43\\
6	47.43\\
8	47.43\\
10	47.43\\
12	47.43\\
14	47.43\\
16	47.43\\
18	47.43\\
20	47.43\\
22	47.43\\
24	47.43\\
26	47.43\\
28	47.43\\
30	47.43\\
32	46.16\\
34	45.64\\
36	50\\
38	60\\
40	70\\
42	80\\
44	90\\
46	100\\
48	100\\
50	100\\
52	100\\
54	100\\
56	100\\
58	100\\
60	100\\
62	100\\
64	100\\
66	100\\
68	100\\
70	100\\
72	100\\
74	100\\
76	100\\
78	100\\
80	100\\
82	100\\
84	100\\
86	100\\
88	100\\
90	100\\
92	100\\
94	100\\
96	100\\
98	100\\
100	100\\
102	100\\
104	100\\
106	100\\
108	100\\
110	100\\
112	100\\
114	100\\
116	100\\
118	100\\
120	100\\
122	100\\
124	100\\
126	100\\
128	100\\
130	100\\
132	100\\
134	99.98\\
136	99.55\\
138	98.87\\
140	98.07\\
142	97.27\\
144	96.55\\
146	95.93\\
148	95.45\\
150	95.11\\
152	94.92\\
154	94.86\\
156	94.84\\
158	94.84\\
160	94.84\\
162	94.84\\
164	94.85\\
166	94.85\\
168	94.86\\
170	94.86\\
172	94.86\\
174	94.86\\
176	94.86\\
178	94.86\\
180	94.87\\
182	94.87\\
184	94.87\\
186	94.86\\
188	94.86\\
190	94.86\\
192	94.86\\
194	94.86\\
196	94.86\\
198	94.86\\
200	94.86\\
202	94.86\\
204	94.86\\
206	94.86\\
208	94.86\\
210	94.86\\
212	94.86\\
214	94.86\\
216	94.86\\
218	94.86\\
220	94.86\\
222	94.86\\
224	94.86\\
226	94.86\\
228	94.86\\
230	94.86\\
232	94.86\\
234	94.86\\
236	94.86\\
238	94.86\\
240	94.86\\
242	94.86\\
244	94.86\\
246	94.86\\
248	94.86\\
250	94.86\\
252	94.86\\
254	94.86\\
256	94.86\\
258	94.86\\
260	94.86\\
262	94.86\\
264	94.86\\
266	94.86\\
268	94.86\\
270	94.86\\
272	94.86\\
274	94.86\\
276	94.86\\
278	94.86\\
280	94.86\\
282	84.91\\
284	74.91\\
286	64.91\\
288	54.91\\
290	44.91\\
292	34.91\\
294	24.91\\
296	14.91\\
298	4.906\\
300	5.28e-05\\
302	2.715e-06\\
304	0.2383\\
306	10.24\\
308	20.24\\
310	30.24\\
312	40.24\\
314	47.13\\
316	49.17\\
318	49.2\\
320	48.64\\
322	48.09\\
324	47.72\\
326	47.51\\
328	47.43\\
330	47.4\\
332	47.41\\
334	47.42\\
336	47.42\\
338	47.43\\
340	47.43\\
342	47.43\\
344	47.43\\
346	47.43\\
348	47.43\\
350	47.43\\
352	47.43\\
354	47.43\\
356	47.43\\
358	47.43\\
360	47.43\\
362	47.43\\
364	47.43\\
366	47.43\\
368	47.43\\
370	47.43\\
372	47.43\\
374	47.43\\
376	47.43\\
378	47.43\\
380	47.43\\
382	47.43\\
384	47.43\\
386	47.43\\
388	47.43\\
390	47.43\\
392	47.43\\
394	47.43\\
396	47.43\\
398	47.43\\
400	47.43\\
};
\end{axis}
\end{tikzpicture}%
}\\
	\subfigure{\input{images/dinput.tikz}}\\
	\caption{Comparison of control performance under authorities of 
	linear-based MPC and non-linear model predictive control.}
	\label{fig:results}
\end{figure}

The performance of proposed control strategies has been tested on a simulation 
scenario involving a single conical tank, described by 
equation~\eqref{eq:tank_H} and parameters reported in the 
table~\ref{tab:params}. We consider a simulation window of \SI{400}{\second}, 
where a reference change, i.e. the desired level of the liquid changes, occurs 
at times $\ui{t}{up} = \SI{50}{\second}$ and at $\ui{t}{down} = 
\SI{350}{\second}$. Specific time profiles of process and manipulated variables 
can be viewed on the Fig.~\ref{fig:results}. 

Both presented approaches have a couple of advantages, which includes 
constraint satisfaction as well as their enforce optimal behavior. Furthermore, 
the nature 
of predictive control can see around the times $\ui{t}{up}$ and 
$\ui{t}{down}$, where the controller reacts in an anticipation of the reference 
changes. Naturally, the non-linear predictive controller handles the changes in 
the level control better, that the linear-based control. The advantage can be 
seen mainly in operation towards lower levels of liquid, there the linear-based 
control undershoots the reference significantly. Note, that such a controller 
cannot be used when considering liquid levels close to the bottom of the tank.

\section{Conclusions}
\label{sec:clc}
This paper covered the design and comparison of two predictive control 
strategies for the most important chemical process, the liquid storage tank. 
Specifically, a conically-shaped storage device was considered. Both 
controllers have enforced constraint satisfaction, which is one of the most 
important tasks in the process control. Moreover, we have shown, that by 
considering a non-linear prediction equation in the controller, we have 
achieved better tracking of the desired liquid level when considering step-down 
reference change. Compared to the linear-based MPC, the NMPC is capable of 
regulating the liquid level even near the bottom of the storage tank.

\section*{Acknowledgments}
The authors gratefully acknowledge the contribution of the Scientific Grant 
Agency of the Slovak Republic  under the grants 1/0403/15, the contribution of 
the Slovak Research and Development Agency under the project APVV 15-0007, and 
the Research \& Development Operational Programme for the project  University 
Scientific Park STU in Bratislava, ITMS 26240220084,  supported by the Research 
7 Development Operational Programme funded by the ERDF. M. Klaučo would like to 
thank for the financial contribution from the STU in Bratislava Grant Scheme 
for Excellent Research Teams.

\section*{References}
\bibliographystyle{apastyle}
\bibliography{klauco_bibfile,klauco_bibfile2}

\begin{thebibliography}{}

\bibitem[Allg{\"o}wer et~al., 2004]{Allgwer2004NonlinearMP}
Allg{\"o}wer, F., Findeisen, R., and Nagy, Z.~K. (2004).
\newblock Nonlinear model predictive control : From theory to application.

\bibitem[Bako\v{s}ov\'a and Oravec, 2014]{uiam1552}
Bako\v{s}ov\'a, M. and Oravec, J. (2014).
\newblock Robust mpc of an unstable chemical reactor using the nominal system
  optimization.
\newblock {\em Acta Chimica Slovaca}, 7(2):87--93.

\bibitem[Camacho and Bordons, 2007]{camacho:2007}
Camacho, E.~F. and Bordons, C. (2007).
\newblock {\em {M}odel {P}redictive {C}ontrol}.
\newblock Springer, 2nd edition.

\bibitem[King, 2010]{king:2010:book}
King, M. (2010).
\newblock {\em Process Control: A Practical Approach}.
\newblock Wiley.

\bibitem[Kvasnica et~al., 2010]{uiam906}
Kvasnica, M., Herceg, M., {\v{C}}irka, {\v{L}}., and Fikar, M. (2010).
\newblock Model predictive control of a cstr: A hybrid modeling approach.
\newblock {\em Chemical papers}, 64(3):301--309.

\bibitem[\L{}awry\'{n}czuk, 2017]{lavrynczuk:2017:isa}
\L{}awry\'{n}czuk, M. (2017).
\newblock Nonlinear predictive control of a boiler-turbine unit: A state-space
  approach with successive on-line model linearisation and quadratic
  optimisation.
\newblock {\em ISA Transactions}, 67:476 -- 495.

\bibitem[Mayne et~al., 2000]{mayne:aut:2000}
Mayne, D.~Q., Rawlings, J.~B., Rao, C.~V., and Scokaert, P. O.~M. (2000).
\newblock Constrained model predictive control: Stability and optimality.
\newblock {\em Automatica}, 36(6):789 -- 814.

\bibitem[Mikle\v{s} and Fikar, 2007]{fikar:2007:book}
Mikle\v{s}, J. and Fikar, M. (2007).
\newblock {\em Process Modelling, Identification, and Control}.
\newblock Springer Verlag, Berlin Heidelberg.

\bibitem[Muske, 1997]{muske:acc:1997:ofs}
Muske, K.~R. (1997).
\newblock Steady-state target optimization in linear model predictive control.
\newblock In {\em American Control Conference, 1997. Proceedings of the 1997},
  volume~6, pages 3597--3601 vol.6.

\bibitem[Muske and Badgwell, 2002]{muske:jpc:2002:ofs}
Muske, K.~R. and Badgwell, T.~A. (2002).
\newblock Disturbance modeling for offset-free linear model predictive control.
\newblock {\em Journal of Process Control}, 12(5):617 -- 632.

\bibitem[Nocedal and Wright, 2006]{nocedal:book:2006}
Nocedal, J. and Wright, S.~J. (2006).
\newblock {\em Numerical Optimization}.
\newblock Springer, New York, 2nd edition.

\bibitem[Rawlings and Mayne, 2009]{rawlings:book:2009}
Rawlings, J.~B. and Mayne, D.~Q. (2009).
\newblock Model predictive control: Theory and design.

\bibitem[Sharma et~al., 2015]{uiam1612}
Sharma, A., Fikar, M., and Bako\v{s}ov\'a, M. (2015).
\newblock Comparative study of time optimal controller with pid controller for
  a continuous stirred tank reactor.
\newblock {\em Acta Chimica Slovaca}, 8(1):27--33.

\end{thebibliography}
\end{document}